# Mountain resonance shapes the distribution of earthquake-induced landslides

Ashok Dahal[1]

[1]University of Twente, Faculty of Geo-Information Science and Earth Observation (ITC), Hallenweg 8, 7522NH, Enschede, Netherlands

Corresponding Author: a.dahal@utwente.nl

## Abstract

Earthquake-induced landslides are among the most destructive cascading hazards, yet the physical mechanisms governing their spatial distribution and size remain incompletely understood. While amplification of seismic waves by local topography is well documented, the role of mountain-scale resonance has remained largely untested. Here we demonstrate that the resonant frequency of mountains exerts a statistically significant control on the occurrence, spatial pattern, and size of coseismic landslides. Using the 2015 Mw 7.8 Gorkha earthquake in Nepal, which triggered more than 25,000 landslides, we combine three-dimensional modal analysis of 3,130 mountain bodies with a high-resolution landslide inventory. Landslides preferentially occur on mountains whose resonant frequencies (~0.5–1.2 Hz) overlap with the dominant frequency content of the earthquake. Mountains affected by landslides exhibit systematically lower resonant frequencies than stable mountains, and larger landslides are associated with lower-frequency mountains. These findings provide quantitative evidence that mountain-scale resonance amplifies seismic ground motion sufficiently to influence landslide triggering, highlighting an overlooked mechanism that should be incorporated into assessments of earthquake-induced landslide hazard.

# 1 Introduction

Earthquake-induced landslides (EQIL) are one of the main cascading hazards among tsunami and liquefaction, caused by an earthquake event[1], [2]. The economic and human cost of EQIL is very significant[3], [4, p. 63]. For example, the 1970 Huascaran earthquake in Peru had an estimated 27,500 landslide fatalities[5], similarly, the 2008 Wenchuan earthquake in China is estimated to have 20,000 fatalities[6]. The geology, slope, and ground motion due to the earthquake are primary controls of the EQIL[2]. Among these, the ground motion is the triggering factor as it perturbs the otherwise stable slope, leading to a failure[7].

The ground motion is a manifestation of the seismic forcing exerted in the given spatial location in the form of a seismic wave, which contains two main properties: the amplitude and frequency. The combined effect of the amplitude and frequency often plays a deciding role in the genesis of landslides[8]. Moreover, the seismic wave can interact with surface, topographic, and subsurface structures to alter the ground motion, which leads to amplification or de-amplification of the ground motion[9], [10], [11], [12], [13]. Many of these interactions and their resulting effect on landslides have been well studied, both in seismology literature and the EQIL literature (see[2]). For example, the topographic amplification has been well studied in terms of seismic wave amplification as well as a factor for landsliding[14], [15], [16], [17]. The soil amplification has been well studied since it was definitively observed in the Mexico City earthquake in 1985[18]. However, the interaction between the seismic wave and the resonance frequency of the mountains is observed but not well established as a factor causing landslides[19].

The natural (undamped) frequency of a system characterises the rate at which it undergoes free vibration after an initial perturbation in the absence of sustained external forcing; it is an intrinsic property set by the system's mass distribution, stiffness characteristics, boundary conditions and geometry[20], [21]. If the dominant frequency content of a seismic input coincides with one of a site's or structure's modal frequencies, resonant amplification can occur, producing substantially larger ground- and structural-motion amplitudes than would arise from non-resonant excitation[14], [22], [23], [24]. This resonance phenomenon is a central consideration in earthquake engineering and underlies seismic design provisions in building codes and guidelines; accordingly, engineers employ modal analysis, response spectra, damping models, and soil–structure interaction analyses to predict, quantify, and mitigate resonance-driven damage[25].

Mountains similarly possess intrinsic vibrational modes, and ambient-vibration[11], [26], [27], [28], and studies have demonstrated measurable resonance frequencies for individual peaks and ridges[19]. For example, a field and simulation-based evaluation by Weber et al.[19] on the Matterhorn reported a dominant resonance frequency of approximately 0.42 Hz. In the same work, authors have postulated that mountains could resonate when the seismic wave matches the resonant frequency of the mountains, leading to a higher probability of landslides due to amplified ground motion. However, this postulation has never been tested, as I need a perfect scenario of an earthquake with the seismic waves in the range of resonance frequencies of the mountain, landslides due to that earthquake, and the resonant frequency for all mountains in observation. Therefore, I do not know if and how the resonant frequency affects landslide genesis and its spatial distribution.

In this research, I hypothesise that the resonant frequency has a statistically significant effect on the genesis, spatial distribution, and size of the landslide at a given spatial domain.

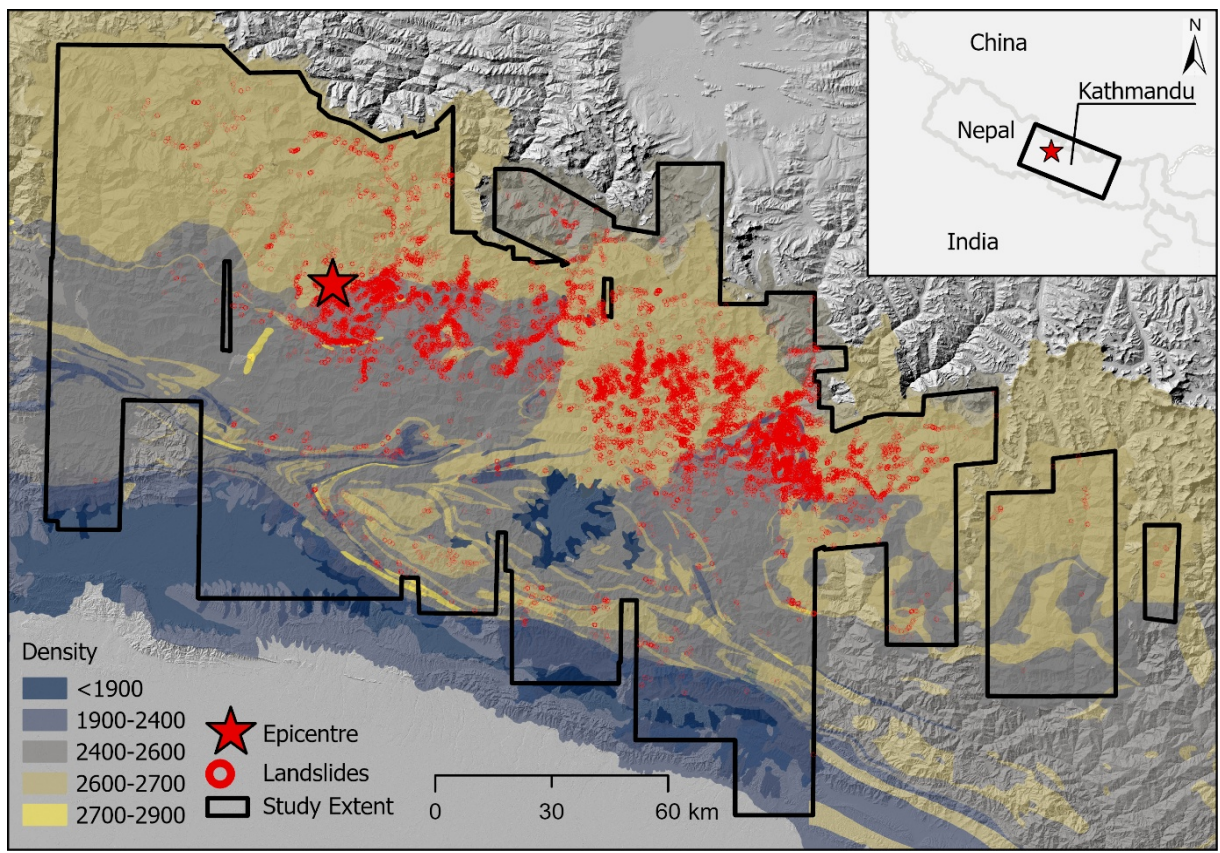

*Figure 1: **Study area and modelling domain defined by the landslide inventory extent.***

## 2 Method

### 2.1 Study Area and Dataset

I evaluated the proposed hypothesis within the region impacted by the 2015 Gorkha Earthquake (Mw 7.8), which triggered more than 25,000 landslides[29]. This study area (Figure 1) was selected primarily because the dominant frequency range of the Gorkha earthquake (<3.0 Hz) overlaps with the indicative resonant frequency range of mountainous terrain, as previously reported[10], [11], [19]. Furthermore, the Nepalese Himalaya exhibits highly variable topography, ranging from gently sloping hills to steep, high-relief mountains, providing an ideal natural laboratory to investigate the interplay between resonant frequency characteristics and landslide occurrence.

The Gorkha earthquake represents one of the largest recorded seismic events along the Main Himalayan Thrust (MHT). This event was characterised by a thrust faulting mechanism typical of the MHT system, with the seismogenic fault exhibiting a strike of 293° and a dip of approximately 7°[30]. According to the USGS (2015), the mainshock nucleated in the northwestern sector of the study area (28.231°N, 84.731°E ± 7.3 km) at a depth of 8.2 ± 2.9 km[31]. The rupture propagated south-eastward along the dip direction, lasting approximately 70 seconds, with a maximum slip of 5.2 m.

Roback et al.[29] produced a detailed coseismic landslide inventory for this event using multi-sensor, very high-resolution optical imagery (0.2–0.5 m). The inventory delineates both source and deposition zones, with landslide areas following a distribution centred at a mean of 3,473 m$^2$ (standard deviation: 11,240 m$^2$). The largest mapped failure, the Langtang Valley landslide, covered an area of approximately 1,720,500 m$^2$. Their analysis revealed that most landslides were concentrated along the east–west trace of the MHT, with density increasing eastward in the direction of rupture propagation. Interestingly, they reported that landslide density was not strongly correlated with peak ground acceleration (PGA) or maximum slope angle. These findings make this inventory particularly valuable for exploring landslide occurrence in relation to the resonance frequencies.

Apart from the landslide and earthquake information, I also used the digital terrain model and geology as the basis for this analysis. The digital terrain model was obtained from the Shuttle Radar Topography mission (SRTM) at a resolution of 30 meters[32]. Even though the DEM itself is at a lower resolution, it can approximately represent the geometric structure of the mountains to estimate the resonant frequency of the mountains that are within the study area. Similarly, the geology dataset that I use is obtained from Dahal[33], which provides the approximate formation of the mountains within the study area, and based on that formation's details and my fieldwork experience, I have estimated the dominant lithological types as well as their average density.

## 2.2 Mountain Delineation and Meshing

With the available SRTM digital elevation model[32], I needed to delineate each mountain as a separate entity to simulate and approximate the resonant frequency of those mountains. For that, I used the standard hydrological approach of catchment delineation, but with an inverse DEM. While using inverse DEM, it ensures to delineate the valleys instead of the ridgelines and provides an approximation of the mountain boundary. Here, I generated the inverse DEM by subtracting the SRTM DEM by 10000 meters. This number is chosen to be high enough to not have any negative elevation, and if it transforms linearly and the slope gradients are preserved, such a number can be selected arbitrarily.

The mountains were delineated using the standard watershed/catchment delineation workflow[34]. Prior to mountain delineation, sinks and depressions were corrected using a fill algorithm to ensure continuous flow paths. Flow direction was computed using the D8 algorithm, which assigns flow from each cell to the steepest downslope neighbour. Subsequently, a flow accumulation grid was generated to represent the number of upstream cells contributing to each location. A threshold of 5000 cells was applied (tested 1000-10000) to the flow accumulation grid to define the stream network automatically. Catchments were then delineated by partitioning the DEM into drainage areas associated with each stream segment, using a minimum catchment size of 1 sq km to exclude very small basins. The defined catchments were then converted into polygons to obtain the mountain boundary (shown in Figure 2a and Figure 2b). Overall, 3130 mountain boundaries are delineated in the study area with a minimum size of 1 sq.km. and a maximum size of 68.74 sq.km. The mean size of the polygons is 9.58 sq.km.

For each delineated mountain, the mesh is constructed to represent the three-dimensional volume at the mountain footprint defined on the DEM. To do so, I cropped the SRTM-DEM to the extent of the delineated mountain and extracted the elevation surface. From the cropped raster, I use the native grid of DEM pixel centres as the horizontal node positions so that the mesh is geographically aligned with the original elevation data; this alignment preserves topographic detail and simplifies subsequent mapping of model outputs back onto the DEM. Horizontal cells that coincide with DEM nodata or otherwise invalid elevations are excluded from the discretisation, so the mesh conforms exactly to the valid terrain footprint and does not contain spurious, isolated elements over missing data.

Vertically, the surface grid is extruded into a small number of uniform layers to produce a stack of trilinear hexahedral elements using Meshio[35]. The top layer nodes are placed at the DEM elevations, and deeper layers are generated by subtracting a fixed layer thickness from the nodes above, with the deepest layer referenced to a consistent lower elevation to avoid degenerate elements where surface data are sparse. This layered extrusion yields structured, well-shaped hexahedra (Figure 2c and d) that follow the surface geometry and concentrate resolution where it matters most for low-frequency modal behaviour: near-surface strata and the upper portion of the mountain mass. The number of vertical layers and the layer thickness are chosen to capture the expected modal wavelengths with minimal degrees of freedom and are refined where necessary based on convergence tests.

Connectivity is created using regular (i, j, k) indexing of the extruded grid, producing straightforward element numbering and efficient assembly for the finite-element solver. Mesh preprocessing includes simple quality checks such as the removal of elements with zero or negative volume, inspection of aspect ratios, and optional local smoothing of node elevations to reduce numerical ill-conditioning. Regions corresponding to external faces (top, bottom and lateral facets) are identified during mesh construction to enable application of physically meaningful boundary conditions during the modal analysis.

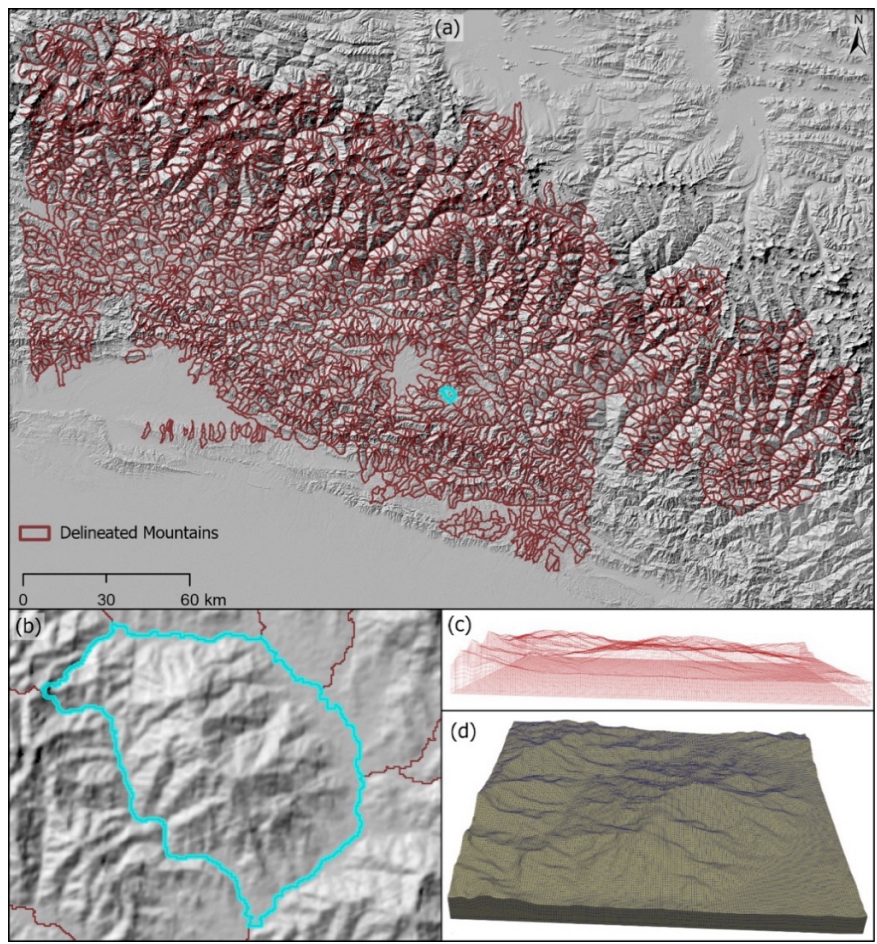

*Figure 2: Study Area and delineated mesh. (a) shows the overall study area with delineated mountain faces. (b) shows a highlighted and zoomed polygon representing a delineated mountain, (c) and (d) represent the mesh for the mountain shown in (b) as a wireframe and hexahedral mesh, respectively.*

## 2.3   Resonance frequency Computation

I model the catchment volume as a three-dimensional, isotropic, linear-elastic continuum and perform an undamped modal (eigenfrequency) analysis on a layered hexahedral finite-element mesh constructed from the digital elevation model (DEM) clipped by the catchment footprint. The overall simulation was performed using SfePY[36] and with the Locally Optimal Block Preconditioned Conjugate Gradient Method (LOBPCG) eigesolver[37].

Let $\mathbf{u}(\mathbf{x}, t) = [u_x, u_y, u_z]^\top$ denote the displacement field at point $\mathbf{x} \in \Omega$ and time $t$. Under the infinitesimal-strain assumption (suitable because the modal analysis linearises about the undeformed reference configuration), the small-strain tensor is the symmetric gradient of displacement,

$$\varepsilon_{ij}(\mathbf{x}, t) = \frac{1}{2}\left(\partial_{x_j} u_i(\mathbf{x}, t) + \partial_{x_i} u_j(\mathbf{x}, t)\right), i, j \in \{1,2,3\}.$$

This linear kinematic assumption is justified for free-vibration analysis of soils and rocks when modal amplitudes are small and when the goal is to identify resonant frequencies and mode shapes rather than to simulate large nonlinear deformations.

For isotropic, linear elasticity, the stress-strain relation (Hooke's law) reads

$$\sigma_{ij} = \lambda \delta_{ij} \varepsilon_{kk} + 2\mu \varepsilon_{ij}$$

with Lamé parameters $\lambda, \mu$ related to Young's modulus $E$ and Poisson's ratio $\nu$ by

$$\mu = \frac{E}{2(1+\nu)}, \lambda = \frac{E\nu}{(1+\nu)(1-2\nu)}$$

Writing stresses and strains in Voigt notation gives the constitutive matrix $\mathbf{C}$ such that $\boldsymbol{\sigma} = \mathbf{C}\boldsymbol{\varepsilon}$. The linear elastic constitutive assumption yields a symmetric positive-definite stiffness operator and captures the first-order elastic response that governs modal behaviour.

The strong form of undamped elastodynamics (neglecting body forces) is

$$\nabla \cdot \boldsymbol{\sigma} + \rho \ddot{\mathbf{u}} = \mathbf{0}$$

where $\rho$ denotes mass density, obtained from the geological data[33]. Multiplying by a virtual displacement $\delta\mathbf{u}$ and integrating over the domain gives the variational statement

$$\int_\Omega \delta\varepsilon : \mathbf{C} : \varepsilon \, d\Omega = \int_\Omega \rho \delta\mathbf{u} \cdot \ddot{\mathbf{u}} \, d\Omega$$

Assuming time-harmonic motion $\mathbf{u}(\mathbf{x}, t) = \mathbf{U}(\mathbf{x})e^{i\omega t}$ with $\ddot{\mathbf{u}} = -\omega^2 \mathbf{U}$ leads to the modal weak form

$$\int_\Omega \delta\varepsilon : \mathbf{C} : \varepsilon \, d\Omega = \omega^2 \int_\Omega \rho \delta\mathbf{u} \cdot \mathbf{u} \, d\Omega$$

which defines the continuous generalised eigenproblem for angular frequency $\omega$. For spatial discretisation, I use trilinear hexahedral (8-node) elements obtained by extruding the DEM-sampled horizontal grid into $N_z$ vertical layers of prescribed thickness. Hexahedral elements are chosen because they align naturally with the regular DEM sampling, reduce the number of elements required for a given resolution compared to tetrahedralizations of an extruded grid, and typically produce better-conditioned stiffness matrices for low-order elasticity problems. Let $\{N_A(\mathbf{x})\}$ denote the scalar trilinear shape functions; the displacement field is approximated as

$$\mathbf{U}(\mathbf{x}) \approx \sum_{A=1}^{n_{\text{node}}} N_A(\mathbf{x}) \mathbf{U}_A,$$

and the strain field in an element is approximated by the strain-displacement relation $\varepsilon(\mathbf{x}) \approx \mathbf{B}(\mathbf{x})\mathbf{d}_e$, where $\mathbf{B}$ is the standard $6 \times 3n_e$ matrix mapping element nodal DOFs $\mathbf{d}_e$, to small strains. The element stiffness and consistent mass matrices follow:

$$\mathbf{K}_e = \int_{\Omega_e} \mathbf{B}^\top(\mathbf{x})\mathbf{C}\mathbf{B}(\mathbf{x})d\Omega, \mathbf{M}_e = \int_{\Omega_e} \rho \mathbf{N}^\top(\mathbf{x})\mathbf{N}(\mathbf{x})d\Omega$$

where $\mathbf{N}$ is the matrix of shape functions assembled for the element, these integrals are evaluated by Gaussian quadrature on the reference hexahedron with a quadrature order sufficient for accurate integration of the bilinear integrands. This choice balances accuracy and computational cost.

Global assembly of elemental contributions yields the algebraic generalised eigenproblem

$$\mathbf{K}\boldsymbol{\Phi} = \lambda \mathbf{M}\boldsymbol{\Phi}, \lambda = \omega^2,$$

where $\boldsymbol{\Phi}$ denotes a global mode shape. Natural frequencies in hertz are $f = \omega/(2\pi) = \sqrt{\lambda}/(2\pi)$. Essential (Dirichlet) boundary conditions are applied by elimination of constrained degrees of freedom; when no constraints are applied, rigid-body modes appear with near-zero eigenvalues. Rather than discarding a fixed number of rigid modes, modes are filtered using a relative threshold $\varepsilon$ (i.e., retaining modes with $\lambda > \varepsilon \max_k \lambda_k$), which is more robust across irregular geometries and avoids accidental truncation of weakly constrained physical modes.

Modal solutions are obtained with an iterative eigensolver capable of returning a small subset of the lowest-frequency eigenpairs[37]. Computed eigenvectors are mass-normalised by

$$m_j = \boldsymbol{\Phi}_j^\top \mathbf{M} \boldsymbol{\Phi}_j, \tilde{\boldsymbol{\Phi}}_j = \boldsymbol{\Phi}_j/\sqrt{m_j},$$

so that $\tilde{\boldsymbol{\Phi}}_j^\top \mathbf{M} \tilde{\boldsymbol{\Phi}}_j = 1$. Numerical validation checks include the eigenpair residual

$$\text{res}_j = \frac{\|\mathbf{K}\boldsymbol{\Phi}_j - \lambda_j \mathbf{M}\boldsymbol{\Phi}_j\|}{\|\lambda_j \mathbf{M}\boldsymbol{\Phi}_j\|},$$

the modal orthogonality condition $\boldsymbol{\Phi}_i^\top \mathbf{M} \boldsymbol{\Phi}_j \approx 0$ for $i \neq j$, and the energy relation $\boldsymbol{\Phi}_j^\top \mathbf{K} \boldsymbol{\Phi}_j = \lambda_j \boldsymbol{\Phi}_j^\top \mathbf{M} \boldsymbol{\Phi}_j$. Mesh convergence studies (varying horizontal resolution and vertical layering) are used to ensure the computed lowest eigenfrequencies are insensitive to discretisation choices.

### 2.4  Understanding the relation with landslides

Once the resonant frequency is identified, I then performed statistical tests to understand if and how the mountains with landslides were separated from the mountains without landslides. For that, I performed two distinct statistical tests, namely Kolmogorov-Smirnov[38] test and the Student's t-test[39].

The Kolmogorov-Smirnov (KS) test is a nonparametric procedure for comparing two distributions (here, resonance frequencies of mountains with landslides and non-landslides) by measuring the maximum vertical distance between empirical cumulative distribution functions (ECDFs). In the one-sample KS test, the null hypothesis is that a sample of size $n$ is drawn from a specified continuous distribution $F$, and the test statistic is

$$D_n = \sup_x |F_n(x) - F(x)|,$$

where, $F_n$ denotes the sample ECDF. In the two-sample KS test, the null hypothesis is that two independent samples of sizes $n$ and $m$ are drawn from the same (unspecified) continuous distribution; the statistic is

$$D_{n,m} = \sup_x |F_n(x) - G_m(x)|,$$

with, $F_n$ and $G_m$ the respective ECDFs. The KS test is distribution-free under the null for continuous $F$, so p-values can be obtained without strong parametric assumptions; it is sensitive to differences in location, scale and shape because it compares full CDFs, but it places emphasis on the largest local discrepancy and thus can be dominated by differences in a limited range.

Similarly, the Student's $t$-test was also used to assess differences in population means of resonance frequencies of mountains with and without landslides. Here, I used the two-sample $t$-test for independent groups, as the sample sizes for landslide and non-landslide groups are different. For two independent samples with means $\bar{x}_1, \bar{x}_2$, variances $s_1^2, s_2^2$ and sizes $n_1, n_2$, Welch's $t$-statistic (which does not assume equal variances) is

$$t = \frac{\bar{x}_1 - \bar{x}_2}{\sqrt{\frac{s_1^2}{n_1} + \frac{s_2^2}{n_2}}},$$

and the null hypothesis is $\mu_1 = \mu_2$. Under the null $t$ is approximately Student-distributed with degrees of freedom given by the Welch–Satterthwaite approximation; the test assumes (approximately) normal sampling distributions and independent observations, and it is generally more powerful than generic distributional tests when the effect of interest is a mean difference.

Apart from the statistical tests, I also examine how the distance from the epicentre, resonant frequency, direction from the epicentre, and the area of the delineated mountain base affect the genesis of landslides. Similarly, I examine the role of resonant frequency on the genesis of landslides with respect to the lithotypes. Apart from variables that can contribute to the landslides, I also examine if and how the size of the landslides themselves is correlated with the resonant frequency of the mountains. This not only tests the hypothesis on the contribution of resonance frequencies to the genesis of landslides but also provides us with an understanding of why landslides can occur at varying resonance frequencies.

## 3 Results

### 3.1 Resonant Frequency Simulation

From modal analysis, I obtained the resonant frequency for each mountain as shown in Figure 3-a. This indicates that the lower part of the study area generally exhibits higher resonant frequencies than the upper part, which is characterised by steep, rocky terrain and snow-covered mountain belts. I also observe in Figure 3-b that the resonant frequency is inversely related to the logarithm of the delineated mountain-base area. In other words, mountains with larger base areas tend to exhibit lower resonant frequencies, and vice versa. This observation indicates that, in general, smaller mountains tend to have higher resonant frequencies, whereas larger mountains tend to have lower resonant frequencies. This behaviour arises because the resonant frequency of a structure depends on the ratio of its stiffness to its mass $\left( f_n \propto \sqrt{k/m} \right)$. Smaller mountains, having less mass and typically greater relative structural stiffness due to their more compact geometry, can resist deformation more effectively and therefore vibrate at higher frequencies[24]. In contrast, larger mountains possess much greater mass and relatively lower overall stiffness, causing them to respond more slowly to dynamic forces and oscillate with lower resonance frequencies[24]. This size–frequency relationship mirrors that observed in engineering structures, such as tall and short buildings.

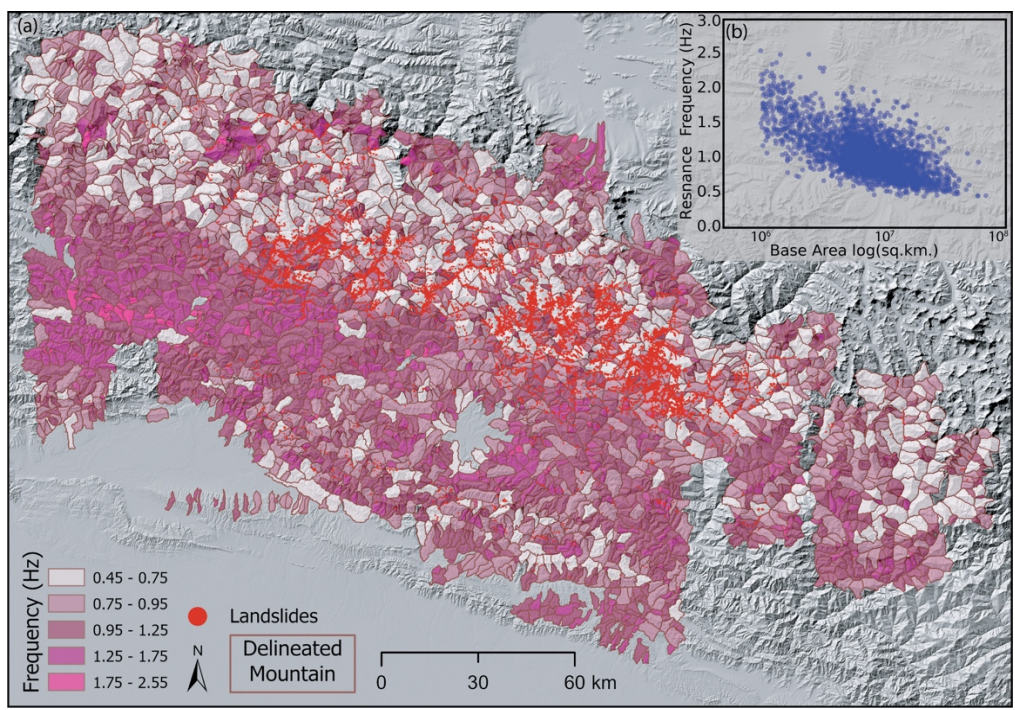

Figure 3: **Results of the modal analysis.** *(a) shows the spatial distribution of the resonant frequency. (b) shows the relation between the resonant frequency and the logarithm of the mountain base area.*

## 3.2 Resonant Frequency Control on Landslide

By overlaying the landslide inventory with the simulated resonant-frequency dataset, I observe in Figure 3-a that most of the mapped landslides (red dots) occur in areas with resonant frequencies between approximately 0.45 and 1.25 Hz. However, I also observe that some locations within the same frequency range do not exhibit landslides, as ground-motion intensity varies spatially and numerous additional factors control landslide occurrence. To test and confirm this observation in a quantitative manner, I evaluated the statistical relationship between resonance frequencies and landslide occurrence. I mainly used two statistical tests, namely: two-sample Kolmogorov–Smirnov (KS)[38] and Welch's t-tests[39]. These tests assess whether the distributions and mean values of resonant frequency differed significantly between landslide-affected and unaffected mountains.

From the KS test, I observed that the mountains with and without landslides during the 2015 Gorkha Earthquake had a test statistic of 0.246 with a p-value of 2.5e-37, showing a statistically significant difference in the resonant frequency between the mountains with landslides and without landslides. The KS test statistic shows the maximum absolute difference between the empirical cumulative distribution function (ECDF) of the sample data and the theoretical cumulative distribution function (CDF), as shown in Figure 4-a. This represents that the distribution of resonant frequency of the mountains with landslides and without landslides has fundamentally different distribution and supports proposed hypothesis.

A similar statistical test, known as Welch's t-test, was used to further assess the difference between the two groups. Unlike the KS test, Welch's t-test not only evaluates whether the distributions differ but also quantifies how far apart the group means are relative to the variability in the data. This test shows the test statistic of -15.15 with a p-value of 4.4e-50, indicating that mountains affected by landslides have substantially lower resonance frequencies than stable mountains without recorded landslides. This can be visualised in Figure 4-b, where I can see that the median and both percentile lines for the distribution without landslides are significantly above those with landslides.

Having my hypotheses validated, I examined how the observed ~25000 landslides and their area relate to the resonant frequency. For each landslide polygon, I identified the resonant-frequency range in which it occurred (Figure 4-c). I observed that most of the landslides occur in the range of 0.5-1.0 Hz. I interpret this as a consequence of the 2015 Gorkha earthquake having dominant energy at similarly low frequencies[40], which likely excited mountain slopes with matching resonance frequencies, triggering widespread failures.

Apart from landslide occurrence, I also analysed landslide size (based on crown area) and found that as landslide size increases, both the mean and standard deviation of the associated resonant frequencies gradually decrease (Figure 4-d). This indicates that larger landslides tend to occur on mountains with lower resonance frequencies. Although the mean resonant frequencies for the largest and smallest 10% of landslides fall within a relatively narrow range (approximately 0.80–0.90 Hz), the trend nonetheless shows a systematic decrease in both the mean and its confidence interval with increasing landslide size.

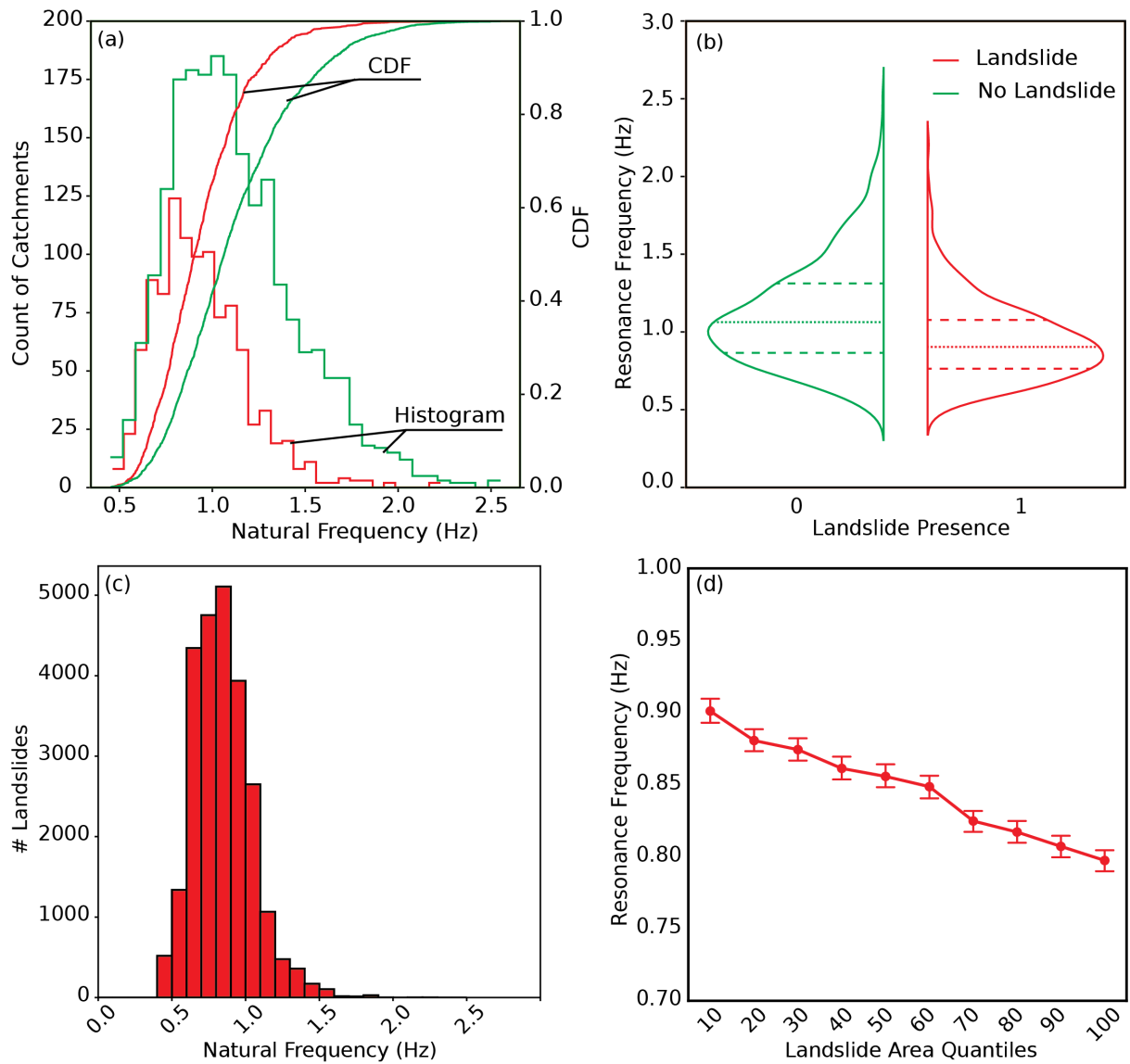

*Figure 4: **Distributions for landslide and non-landslide locations.** (a) ECDF of landslide and non-landslide locations (b) resonance frequency distribution for landslide and non-landslide locations (c) overall distribution of landslide numbers w.r.t. resonance frequency (d) landslide area and their relation with resonance frequency.*

### 3.3 Spatial Control of Landslides w.r.t Earthquake

To understand how the resonant frequency interplays with earthquake in terms of the spatial distribution of landslides, I performed additional analysis. This analysis provided more insights into how the mountains with and without landslides are spatially distributed with respect to the earthquake epicentre.

First, I examine how mountain size and distance from the earthquake epicentre influence the occurrence of landslides, in combination with their resonant frequency. Because neither mountain area nor epicentral distance directly determined landslide size [41] I did not evaluate landslide area with respect to those variables. To visualise these relationships, I plotted the three variables in Figure 5-a: the angular coordinate represents the arctangent of the logarithmic ratio between mountain area and distance from the epicentre, and the radial axis corresponds to resonant frequency. The resulting distribution shows that, for this earthquake, landslides predominantly occurred in mountains that were closer to the epicentre, had relatively larger areas, and possessed resonance frequencies concentrated in the 0.5–1.2 Hz range (also shown in Figure 5-c and Figure 5-d). This cluster suggests that the combined effect of proximity to the rupture source and larger mountain mass, which typically lowers the resonant frequency, created conditions favourable for resonance with the dominant frequency content of the seismic waves. Interestingly, some smaller mountains near the epicentre exhibit higher resonance frequencies and have the presence of landslides (Figure 5-d, 0-50km). However, for increasing distances, the highest frequency slowly starts to decrease.

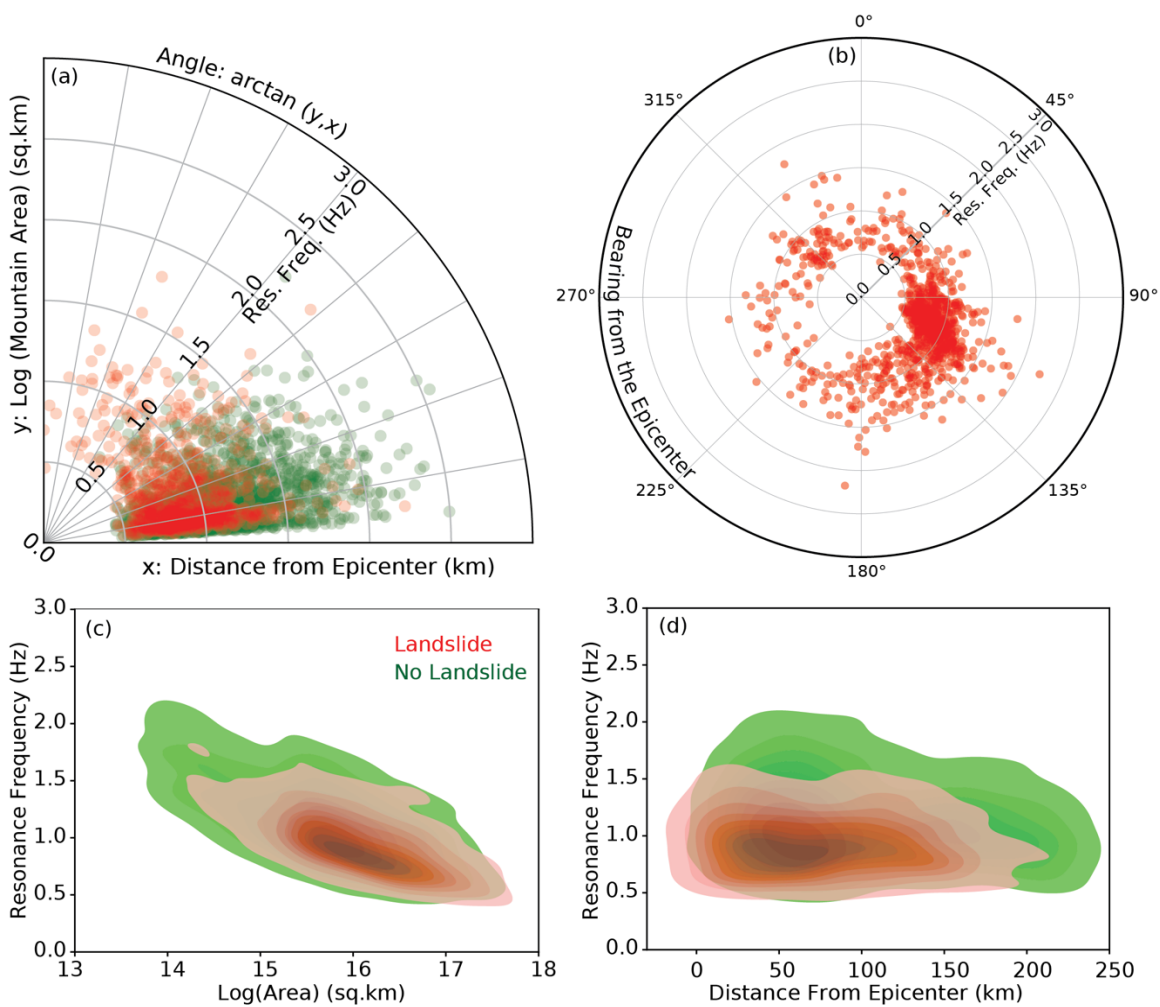

*Figure 5: **Landslide Control of Distance and Direction from Earthquake Epicentre.** (c), (d), and (a) shows the relation with respect to distance and mountain size and their combination, respectively. Similarly, (b) shows the control of direction from the epicentre on landslide genesis.*

Similarly, when examining the directivity effect of the observed landslides (Figure 5-b), I found that in the bearing of 90-180 degrees, the landslides have a higher resonant frequency. The distribution of resonance frequencies associated with landslide-affected mountains remains broadly consistent regardless of direction in the lower frequency band (<1.0Hz), indicating that landslides occurred within a similar frequency range throughout the study area for lower resonant frequency. However, for the higher frequency band (>1.0Hz), I observe that the bearing of 90-180 degrees has a very high cluster of landslides.

## 4 Discussion

In this research, I validate the hypothesis[14], [24], [19] that the resonant frequency of the mountain has an influence on landslide genesis due to the ground motion amplification by topographic resonance. Tested on the 2015 Gorkha earthquake, I show that the low-frequency seismic waves[40] were amplified by mountains with low-resonance frequency, leading to coseismic landslides. However, I must also acknowledge that the resonance does not affect the landslide alone as seismic wave interacts with the top-surface topography and soil structures, creating topographic amplifications, soil (de-) amplifications, etc., and all of that has a combined effect on the genesis of landslides.

The observation on ground motion amplification due to resonance is not new; more than 10 times spectral amplification was observed in the Matterhorn[19]. Similarly, for a hill in Greece with ~0.7Hz resonant frequency, 2-3 times amplification was observed[19], and past studies have indicated that landforms typically exhibit peak spectral amplification factors ranging between 3 and 5, corresponding to resonant frequencies of about 1–3 Hz[27]. However, the finding that it has a statistically significant role in the genesis of landslides is new.

The past attempts to model the genesis of coseismic landslides via Newmark's approach and data-driven method showed some mismatch in the locations where topographic amplification could generate ladnsldies[42]. In that study, the authors did not account for the amplification due to the resonance, which could have certainly exceeded the critical threshold, leading to a slope failure. The primary assumption behind Newmark's approach is that cumulative ground motion which exceeds critical acceleration causes deformation leading to slope failure. If I cannot accurately represent the amplified ground motion due to the resonance of the mountain, I cannot effectively estimate the leading deformation and therefore the landslides. In locations which are very near the epicentre, this does not matter a lot because the ground motion is already very strong, but for locations where lower-frequency seismic waves interact, the effect could be significant[42]. This implies a need to accurately incorporate amplification due to resonance in coseismic landslide models.

Furthermore, the observation of a higher number of landslides in the lower frequency range throughout the study area (with enough ground motion obviously) is because the lower-frequency wavelengths do not get attenuated within the nearfield (<50km)[7]. However, the higher-frequency waves get attenuated quickly and lose their energy, thus disappearing after a certain distance. With the landslides observations (Figure 4-d), I can see that the contour line for landslides has a downward slope, meaning that while going from lower to higher distance, the mountains with higher resonance frequency do not resonate enough to cause a landslide.

Another important point of discussion is the relationship between landslide size and the resonance frequency of the mountains. I observed that the confidence intervals of landslide size remain similar across all quantiles, while the overall trend decreases with increasing resonant frequency, indicating that lower frequencies generally excite larger landslides. As discussed before, for tall or large mountains, amplification can reach up to tenfold, substantially increasing the resulting ground shaking[19]. Similarly, Dunham et al.[41] showed that ground shaking, when combined with factors such as slope and lithology, exerted a significant control on landslide size for coseismic landslides in the Gorkha earthquake, which represents a direct correlation between the amplification due to resonance and the landslide size.

In terms of directivity, I did not find any substantial evidence of directivity and resonance frequency interactions causing landslides in the lower frequency range, as the low-frequency waves are generated and propagated in all directions (even with lower amplitude). However, in the higher frequency range, the landslide pattern has a pronounced directivity effect, which is well expected as the resonance/amplification is caused by the frequency of the incoming wave and often due to the fault geometry, as waves propagate with distinct directivity. The cluster of landslides with higher frequency has a similar direction to that of the major wave propagation direction of 135 degrees[7]. This is consistent with the observation of directivity effects at Matterhorn, showing the direction of incoming wave influences the amplification[19].

There are some limitations in the study I present here; the main limitation is that the available geotechnical/geological data is not of high (spatial) resolution. Since this study focuses on comparing mountains (and their landslides) with similar simulation criteria, it is a valid approach. However, to identify a very accurate resonant frequency, a more detailed dataset would be required. Similarly, due to the lack of seismic stations at the top and bottom of the mountains during the earthquake, I could not observe the resonance and amplification due to the earthquake. Furthermore, a more detailed analysis with actual modelling of slope failure under resonance conditions is recommended in future work to further mechanically validate this observation.

## 5   Conclusion

In conclusion, in this work, I quantitatively prove that the resonant frequency and the subsequent resonance of the mountain play a statistically significant role in causing earthquake-induced landslides. Specifically, for the case of the Gorkha Earthquake, I observe that the mountains with a lower resonant frequency caused more landslides, also with a larger size. Where the resonance effect can cause a landslide is dependent on the frequency of the seismic wave, the directivity of the ground motion and the shape of the mountain itself. This finding opens a new opportunity to investigate the physical role resonance resonating frequency plays in causing landslides, both mechanically as well as due to the ground motion amplification.

### Data Availability

All the used third-party data is available openly through the relevant references in manuscript[29], [32], [33]. The produced dataset within this research is available via Zenodo repository https://doi.org/10.5281/zenodo.18267876.

### Code Availability

All the code necessary to reproduce the results and plots in this research can be found at https://doi.org/10.5281/zenodo.18267876. All the plotting and analysis libraries which are needed to run the code can be installed through openly accessible Pypi repository of python. For any issues/bugs related to running the code please contact the corresponding author via email.